\documentclass[%
 reprint,
 amsmath,amssymb,
 aps,
]{revtex4-2}
\usepackage{mathptmx}
\usepackage{etoolbox}
\usepackage{xcolor}
\usepackage{booktabs}
\usepackage{float}
\usepackage{hyperref}
\usepackage{multirow}
\usepackage{graphicx}
\usepackage{dcolumn}
\usepackage{bm}

\begin{document}

\preprint{APS/123-QED}

\title{Two-Dimensional Hexagonal $\text{BX}_3$ ($\text{X} = \text{P, As}$) Monolayers as High-Capacity, Fast-Charging Anode Platforms for Lithium- and Sodium-Ion Batteries}

\author{Jakkapat Seeyangnok$^{1}$}
 \email{jakkapatjtp@gmail.com} 
 \author{Joongjai Panpranot$^{2,3}$}%
 \email{joongjai.p@chula.ac.th}
\author{Udomsilp Pinsook$^{1}$}%
 \email{Udomsilp.P@Chula.ac.th}
 
\affiliation{$^{1}$Department of Physics, Faculty of Science, Chulalongkorn University, Bangkok 10330, Thailand.}

\affiliation{$^{2}$CrystalLyte Co., Ltd., Research Unit 904, Faculty of Engineering, Chulalongkorn University, Bangkok, 10330, Thailand.}

\affiliation{$^{3}$Center of Excellence on Catalysis and Catalytic Reaction Engineering, Department of Chemical Engineering, Faculty of Engineering, Chulalongkorn University, Bangkok 10330 Thailand.}

\date{\today}

\begin{abstract}
The rapid advancement of alkali-metal ion batteries demands robust anode platforms combining high specific capacities with rapid charge-discharge kinetics. Using first-principles density functional theory (DFT), we systematically evaluate two-dimensional (2D) hexagonal $\text{BX}_3$ ($\text{X} = \text{P}, \text{As}$) monolayers as high-performance dual-use anodes for lithium-ion (LIBs) and sodium-ion batteries (SIBs). Both metallic host architectures display strong thermodynamic affinities for $\text{Li}^+$ and $\text{Na}^+$ adsorption, favoring the hollow $\text{H}_3$ site through synergistic ionic charge transfer and orbital hybridization. Climbing image nudged elastic band (CI-NEB) calculations reveal low direct $\text{H}_3 \rightarrow \text{H}_3$ diffusion barriers: $0.40\text{ eV}$ ($\text{BP}_3$) and $0.26\text{ eV}$ ($\text{BAs}_3$) for $\text{Li}^+$, and $0.26\text{ eV}$ ($\text{BP}_3$) and $0.19\text{ eV}$ ($\text{BAs}_3$) for $\text{Na}^+$, confirming exceptional high-rate kinetics. Thermodynamic convex hulls establish maximum stable lithiation at $\text{Li}_3\text{BX}_3$, yielding low average operating potentials of $0.39\text{ V}$ ($\text{BP}_3$) and $0.35\text{ V}$ ($\text{BAs}_3$) alongside theoretical specific capacities of $775\text{ mAh/g}$ and $341\text{ mAh/g}$, respectively, with $\text{BP}_3$ doubling commercial graphite ($372\text{ mAh/g}$). For SIBs, multi-layer sodiation expands storage up to $\text{Na}_{15}\text{BP}_3$ and $\text{Na}_{12}\text{BAs}_3$, delivering ultrahigh capacities of $3875\text{ mAh/g}$ ($\text{BP}_3$) and $1365\text{ mAh/g}$ ($\text{BAs}_3$) at low voltages of $0.18\text{ V}$ and $0.15\text{ V}$. Crucially, projected density of states (PDOS) analyses confirm that both frameworks preserve intrinsic metallic conductivity throughout all charging stages. These combined properties establish 2D $\text{BX}_3$ monolayers as outstanding, structurally resilient anode candidates for next-generation LIB and SIB energy storage technologies.
\end{abstract}

\keywords{Two-dimensional materials; Alkali-metal ion batteries; Density functional theory; Anode materials; Rapid diffusion; Electronic structure}

\maketitle

\section{Introduction}
Rechargeable batteries utilizing alkali metal ions have emerged as the paramount energy storage technology for modern electronics and grid-scale networks \cite{goodenough2013li,goodenough2010challenges}. Lithium-ion batteries (LIBs) form the bedrock of this infrastructure owing to their high working potential, low self-discharge rates, and superior energy density \cite{li201830,tarascon2001issues,noll2018libs}. Despite these strengths, conventional LIBs face severe bottlenecks, including lithium resource scarcity, rising costs, transition metal extraction concerns \cite{larcher2015towards}, and thermal runaway risks exacerbated by flammable electrolytes \cite{feng2018thermal,kim2019lithium,grosjean2012assessment,liu2014high}. Furthermore, the specific capacity ceiling of commercial graphitic anodes ($372\text{ mAh/g}$ for $\text{Li}^+$) \cite{nitta2015li} restricts their ability to satisfy massive grid storage demands. Consequently, engineering alternative anode platforms remains a critical imperative \cite{sharma2023progress}.

To circumvent these challenges, sodium-ion batteries (SIBs) have emerged as a highly promising, resource-abundant alternative that readily adapts established LIB design principles \cite{hwang2017sodium,pan2013room}. While the larger radius and heavier mass of the $\text{Na}^+$ ion pose kinetic challenges regarding solid-state diffusion, SIBs remain exceptionally viable for large-scale storage where cost-efficiency takes precedence over volumetric density \cite{kim2016recent}. Compared to emerging all-solid-state batteries \cite{wang2024niobium,wang2025high}, SIBs benefit from simpler manufacturing routes and reduced material costs while maintaining highly competitive electrochemical performance.

The realization of high-performance alkali-metal ion batteries is dictated by the choice of anode material, which governs capacity, rate capability, and cycling stability \cite{perveen2020prospects}. An ideal anode must facilitate rapid ion diffusion, preserve structural integrity, maintain electrical conductivity, and operate at low working voltages \cite{li2018recent}. Traditional bulk anodes such as graphite, hard carbon, and metal oxides frequently suffer from sluggish kinetics, mechanical degradation, and severe volume expansion \cite{qiao2023advanced}. Consequently, research has pivoted toward two-dimensional (2D) nanomaterials. Atomically thin 2D sheets provide exceptional surface-to-volume ratios, flexible architectures, and abundant active adsorption sites that boost theoretical storage capacity while ensuring superior reversibility and rapid diffusion \cite{mao2018two,yuan2023review}.

Extensive density functional theory (DFT) screening has evaluated several distinct 2D families. Among elemental monolayers, pristine graphene delivers a double-sided lithium adsorption capacity up to $744\text{ mAh/g}$ \cite{sato1994mechanism}, while phosphorene achieves $432.79\text{ mAh/g}$ for lithium storage with a remarkably low diffusion barrier of $0.09\text{ eV}$ \cite{zhao2014potential}. Silicene offers lithium intercalation approaching bulk silicon ($4200\text{ mAh/g}$), albeit hindered by severe structural strain \cite{tritsaris2013adsorption}. Transition metal dichalcogenides (TMDs) like $\text{MoS}_2$ push lithium capacities beyond $600\text{ mAh/g}$ \cite{stephenson2013lithium}, though structural transitions between their $1\text{T}'$ and $2\text{H}$ phases dramatically influence stability, and their poor intrinsic conductivity often requires carbonaceous hybridization \cite{acerce2015metallic}. Transition metal carbides and nitrides (MXenes), such as metallic $\text{V}_2\text{N}$, achieve exceptional lithium capacities ($925\text{ mAh/g}$) alongside near-zero diffusion barriers \cite{liu2022two}. However, their surface functional terminations ($-\text{F}$, $-\text{OH}$) frequently obstruct transport, necessitating pristine or selectively functionalized surfaces for optimal performance \cite{tang2012mxenes,anasori20172d}.

Amidst these candidates, boron-based nanomaterials have garnered intense interest. Elemental borophene demonstrates exceptional theoretical capacities for lithium storage ($1239$ to $1860\text{ mAh/g}$) \cite{zhang2016could,jiang2016borophene}, while carbon-hybridized $\text{BC}_3$ provides highly favorable adsorption sites for multiple alkali metal ions ($\text{Li}$, $\text{Na}$, $\text{K}$) \cite{joshi2015hexagonal}. Extending this paradigm to group III-V networks, boron pnictides demonstrate exceptional promise. Boron phosphide (BP) and its derivatives ($\text{BP}_2$, $\text{B}_3\text{P}$) undergo favorable semiconductor-to-metal transitions upon ion adsorption, ensuring robust conductivity, delivering capacities up to $1691\text{ mAh/g}$ for both $\text{Li}$ and $\text{Na}$ storage, and providing ultralow diffusion barriers \cite{jiang2017boron,abbas2020two,ye2021metallic}. Two-dimensional boron arsenide ($h$-BAs) possesses excellent thermodynamic stability, delivering capacities of $522\text{ mAh/g}$ for both $\text{Li}$ and $\text{Na}$ ions, alongside low operating voltages \cite{khossossi2019ab}. Furthermore, BP and $h$-BAs monolayers function as highly efficient anchoring matrices in lithium-sulfur (Li-S) configurations, successfully mitigating the polysulfide shuttle effect while improving electronic conductivity \cite{yu2019boron,khossossi2020rational}.

Building upon these established advantages, this work presents a comprehensive first-principles evaluation of $\text{BX}_3$ ($\text{X} = \text{P}, \text{As}$) monolayers as high-performance dual-ion anodes. Following recent computational investigations of the $\text{BP}_3$ framework for Na-ion storage \cite{vu2026bp} and the validation of intrinsic phase stabilities for both $\text{BP}_3$ and $\text{BAs}_3$ via density functional perturbation theory (DFPT) \cite{bp3_seeyangnok,bas3_seeyangnok}, we significantly extend their functional scope. Utilizing DFT, we systematically map their single- and multi-layer alkali metal ($\text{Li}$ and $\text{Na}$) adsorption configurations, multi-stage thermodynamic stabilities, and electrochemical voltage profiles. By rigorously evaluating the underlying charge transfer dynamics, localized chemical bonding, and surface migration trajectories, we detail the evolution of their electronic structures during sequential charging. Ultimately, we demonstrate that both $\text{BP}_3$ and $\text{BAs}_3$ monolayers preserve an intrinsically metallic architecture across their entire operational ranges, establishing them as highly promising, structurally resilient, and fast-charging 2D platforms for next-generation, high-capacity energy storage systems.

\section{Computational Methods}
First-principles calculations were performed within the framework of density functional theory (DFT) as implemented in the \textsc{Quantum ESPRESSO} package~\cite{giannozzi2009quantum}. Electron exchange--correlation effects were treated using the generalized gradient approximation (GGA) parameterized by the Perdew--Burke--Ernzerhof (PBE) functional~\cite{perdew1996generalized}. The interactions between valence electrons and ionic cores were modeled using optimized norm-conserving Vanderbilt (ONCV) pseudopotentials~\cite{hamann2013optimized,schlipf2015optimization}. Throughout the simulations, a plane-wave kinetic-energy cutoff of 80~Ry and a charge-density cutoff of 320~Ry were applied. For the integration of the Brillouin zone, a $12\times12\times1$ Monkhorst--Pack $\mathbf{k}$-point mesh~\cite{monkhorst1976special} was utilized in combination with a Methfessel--Paxton smearing technique~\cite{methfessel1989high} featuring a width of 0.02~Ry.

To assess the electrochemical performance and phase stability of the proposed $\text{BX}_3$ ($\text{X} = \text{P}, \text{As}$) monolayers during alkali-metal intercalation, we evaluated their formation energy landscapes and open-circuit voltage (OCV) profiles. The thermodynamic stability of the intermediate intercalated phases was investigated by calculating the total formation energy ($E_f$) as a function of ion concentration ($x$):
\begin{equation}
E_f = E_{\text{tot}}(\text{M}_x\text{BX}_3) - E{\text{tot}}(\text{BX}_3) - x \cdot E_{\text{tot}}(\text{M}_{\text{bulk}}),
\end{equation}
where $\text{M}$ denotes the intercalated alkali metal ($\text{M} = \text{Li}, \text{Na}$), and $x$ is defined as the adatom concentration ratio relative to the pnictogen atoms ($x = N_{\text{adatom}} / N_{\text{X}}$, with $N_{\text{X}} = 8$ for the host supercell). Here, $E_{\text{tot}}(\text{M}_x\text{BX}_3)$, $E_{\text{tot}}(\text{BX}_3)$, and $E_{\text{tot}}(\text{M}_{\text{bulk}})$ represent the total energies of the intercalated structure, the pristine monolayer, and the corresponding bulk metallic phase (bcc $\text{Li}$ or bcc $\text{Na}$), respectively. To determine the stable intermediate phases, a lower convex hull envelope was constructed across the $E_f$ versus $x$ space. Configurations residing directly on this hull are thermodynamically stable, whereas those lying above it are metastable or unstable, meaning they will spontaneously undergo phase separation into the two closest stable compositions.

The OCV profile as a function of lithium content was subsequently derived from these stable hull phases. For a lithiation step proceeding between two adjacent stable concentrations, $x_{i-1}$ and $x_i$, the average operating voltage ($V$) relative to $\text{Li/Li}^+$ is calculated as:
\begin{equation}
V = -\frac{E_f(x_i) - E_f(x_{i-1})}{(x_i - x_{i-1})e}.
\end{equation}
This thermodynamic approach yields a characteristic step-like voltage profile, where each plateau corresponds to a two-phase coexistence region during the transition between sequential stable lithiated phases.

\section{Results and Discussion}
\subsection{Electrochemical Storage Capacity, Voltage Profiles, and Structural Intercalation Pathways}
    \begin{figure}[ht]
        \centering
        \includegraphics[width=8.7cm]{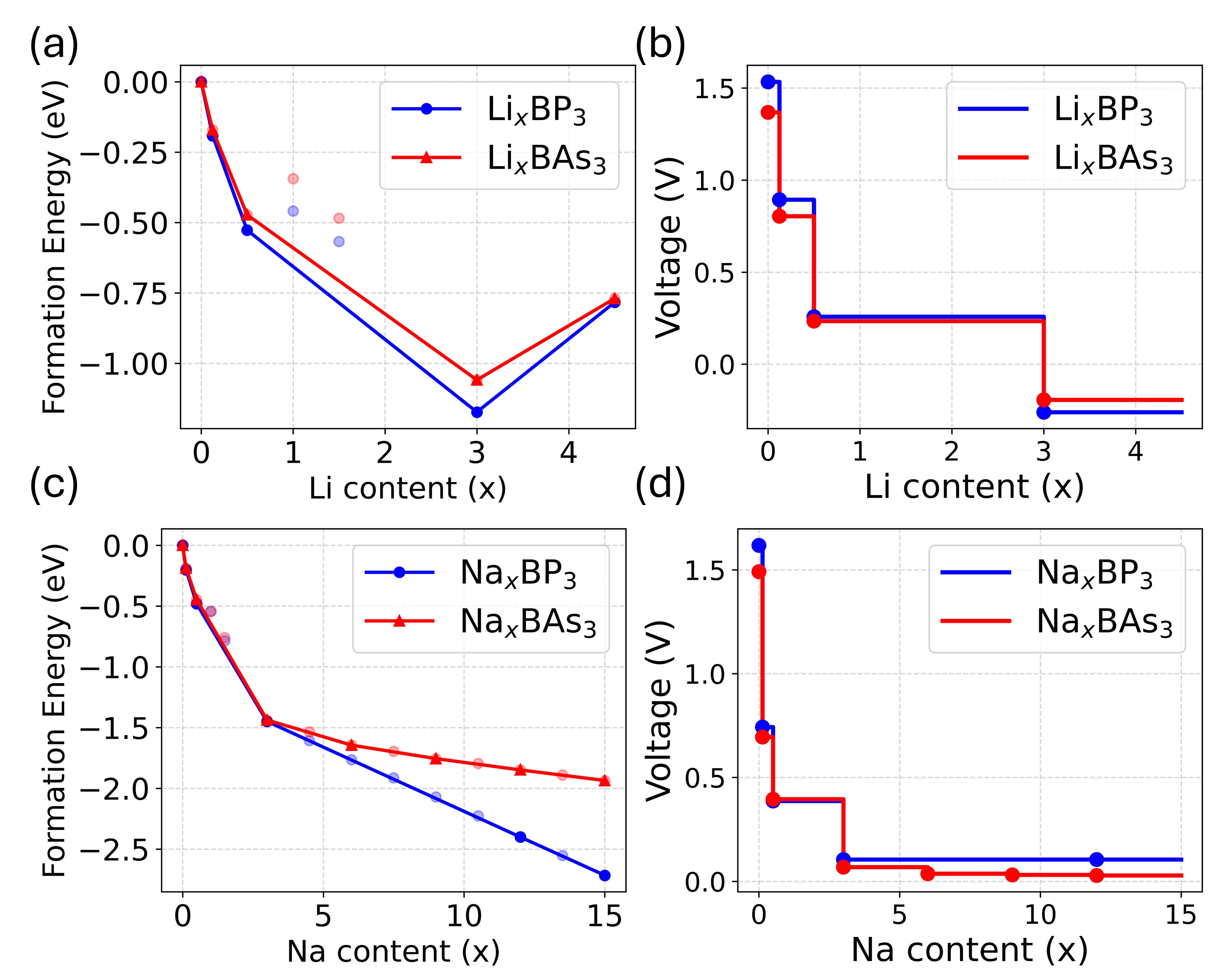}
        \caption{Thermodynamic stability and electrochemical voltage characteristics of $\text{BP}_3$ (blue circles) and $\text{BAs}_3$ (red triangles) monolayers during sequential lithiation and sodiation. (a, c) Derived formation energy convex hulls as a function of alkali concentration $x$ ($x = N_{\text{adatom}}/N_{\text{X}}$, where $N_{\text{X}} = 8$) for (a) lithium insertion ($\text{Li}_x\text{BX}_3$) up to $x = 4.50$ and (c) sodium insertion ($\text{Na}_x\text{BX}_3$) up to $x = 15.00$. Solid lines connect thermodynamically stable ground-state phases, while faintly shaded points indicate metastable or unstable configurations. (b, d) Corresponding stepwise open-circuit voltage (OCV) profiles for (b) lithiation relative to $\text{Li/Li}^+$ and (d) sodiation relative to $\text{Na/Na}^+$ across their full intercalation ranges.}
        \label{fig:ocv_profile}
    \end{figure}

\begin{table*}[htbp]
\centering
\caption{Calculated electrochemical parameters, including average adsorption energy ($E_{\mathrm{ad}}$), total formation energy (Total $E_{\mathrm{ad}}$), and open-circuit voltage (OCV), of $\mathrm{Li}$- and $\mathrm{Na}$-adsorbed $\mathrm{BP}_3$ and $\mathrm{BAs}_3$ monolayers at varying ion concentrations ($x$).}
\label{tab:electrochemical_properties}
\small
\begin{tabular}{cc ccc ccc}
\hline\hline
\toprule
\multirow{2}{*}{Ions} & \multirow{2}{*}{$x$} & \multicolumn{3}{c}{$\mathrm{BP}_3$} & \multicolumn{3}{c}{$\mathrm{BAs}_3$} \\
\cmidrule(lr){3-5} \cmidrule(lr){6-8}
& & $E_{\mathrm{ad}}$ (eV) & Total $E_{\mathrm{ad}}$ (eV) & OCV (V) & $E_{\mathrm{ad}}$ (eV) & Total $E_{\mathrm{ad}}$ (eV) & OCV (V) \\
\hline
\multirow{7}{*}{Lithium (Li)}
& 0.000 &  0.00 &  0.00 &  1.53 &  0.00 &  0.00 &  1.37 \\
& 0.125 & -1.53 & -0.19 &  1.53 & -1.37 & -0.17 &  1.37 \\
& 0.500 & -1.05 & -0.53 &  0.89 & -0.95 & -0.47 &  0.80 \\
& 1.000 & -0.46 & -0.46 &  0.26 & -0.34 & -0.34 &  0.23 \\
& 1.500 & -0.38 & -0.57 &  0.26 & -0.32 & -0.48 &  0.23 \\
& 3.000 & -0.39 & -1.17 &  0.26 & -0.35 & -1.06 &  0.23 \\
& 4.500 & -0.17 & -0.78 & -0.26 & -0.17 & -0.77 & -0.19 \\
\hline
\multirow{14}{*}{Sodium (Na)}
& 0.000 &  0.00 &  0.00 &  1.62 &  0.00 &  0.00 &  1.49 \\
& 0.125 & -1.62 & -0.20 &  1.62 & -1.49 & -0.19 &  1.49 \\
& 0.500 & -0.96 & -0.48 &  0.74 & -0.89 & -0.45 &  0.70 \\
& 1.000 & -0.54 & -0.54 &  0.39 & -0.54 & -0.54 &  0.40 \\
& 1.500 & -0.52 & -0.78 &  0.39 & -0.51 & -0.77 &  0.40 \\
& 3.000 & -0.48 & -1.44 &  0.39 & -0.48 & -1.44 &  0.40 \\
& 4.500 & -0.36 & -1.62 &  0.11 & -0.34 & -1.53 &  0.07 \\
& 6.000 & -0.29 & -1.74 &  0.11 & -0.27 & -1.62 &  0.07 \\
& 7.500 & -0.26 & -1.95 &  0.10 & -0.23 & -1.73 &  0.04 \\
& 9.000 & -0.23 & -2.07 &  0.10 & -0.20 & -1.80 &  0.04 \\
& 10.500 & -0.21 & -2.21 &  0.10 & -0.17 & -1.79 &  0.03 \\
& 12.000 & -0.20 & -2.40 &  0.10 & -0.15 & -1.80 &  0.03 \\
& 13.500 & -0.19 & -2.57 &  0.10 & -0.14 & -1.89 &  0.02 \\
& 15.000 & -0.18 & -2.70 &  0.10 & -0.13 & -1.95 &  0.02 \\
\bottomrule
\hline\hline
\end{tabular}
\end{table*}

    To evaluate the capability and feasibility of the proposed BP$_3$ and BAs$_3$ monolayers as high-capacity dual-ion anodes for next-generation batteries, we systematically computed the stable intermediate phases during progressive Li and Na loading. The calculated thermodynamic convex hulls based on total formation energies, alongside the corresponding open-circuit voltage (OCV) profiles, are illustrated in Fig.~\ref{fig:ocv_profile}, with key metrics compiled in Table~\ref{tab:electrochemical_properties}.

    The thermodynamic stability of the intercalation processes was tracked by constructing formation energy hulls (Fig.~\ref{fig:ocv_profile}a for lithiation, Fig.~\ref{fig:ocv_profile}c for sodiation). The concentration $x$ is defined as the ratio of adsorbed alkali atoms to pnictogen atoms ($x = N_{\text{adatom}} / N_{\text{X}}$), where $N_{\text{X}} = 8$ represents the total number of $\text{P}$ or $\text{As}$ atoms in the host supercell. During Li insertion, the total adsorption energy remains steadily negative down to a concentration of $x = 3.0$. The appearance of clear cusps at $x = 0.5$ and $x = 3.0$ establishes these states as thermodynamically stable configurations against phase separation (Fig.~\ref{fig:ocv_profile}a). However, increasing the Li content further to $x = 4.5$ causes an upward turn in the convex hull curve (yielding total adsorption values of -0.78 eV for BP$_3$ and -0.77 eV for BAs$_3$), indicating that over-lithiation beyond $x = 3.0$ is thermodynamically unfavorable. Conversely, the adsorption of Na ions remains an exothermic and highly favorable process across a much wider concentration range up to $x = 15.0$ (Fig.~\ref{fig:ocv_profile}c). Configurations lying precisely on the solid lines of the convex hull represent the stable intermediate phases. As sodiation progresses, BP$_3$ demonstrates increasingly more negative formation energies compared to BAs$_3$, suggesting a stronger binding affinity for Na atoms and inherently greater thermodynamic stability in heavily sodiated states. These threshold behaviors are clearly confirmed by the step-wise OCV profiles presented in Fig.~\ref{fig:ocv_profile}b (Li) and Fig.~\ref{fig:ocv_profile}d (Na). Atomistically, these macroscopic voltage steps reflect the progressive and spatially distributed occupation of the host lattice. 
    
    During lithiation, the monolayers present initial voltage plateaus at 1.53 V for $\text{BP}_3$ and 1.37 V for $\text{BAs}_3$ ($x = 0.125$), driven by the exceptionally strong binding affinity of Li atoms at dilute concentrations. As the Li framework concentrates toward $x = 1.00$, the system must balance local binding energy against the increasing electrostatic repulsion generated by adjacent $\text{Li}^+$ ions. To minimize this inter-ionic penalty, the adatoms adopt a spatially distributed arrangement across the host surface, which manifests electrochemically as a prolonged, flat OCV plateau of approximately 0.26 V for $\text{BP}_3$ and 0.24 V for $\text{BAs}_3$ spanning the capacity interval from $x = 1.00$ to $x = 3.00$. 
    
    The abrupt voltage drop observed beyond $x = 3.00$ is directly correlated with a critical transition in the storage mechanism. Once the primary surface capacity is fully saturated, subsequent Li insertion is forced into less energetically favorable configurations (Table \ref{tab:electrochemical_properties}). The high energetic cost of over-lithiation is evidenced by a noticeable reduction in average adsorption energy ($E_{\mathrm{ad}}$) from -0.39 eV to -0.17 eV for $\text{BP}_3$ and from -0.35 eV to -0.17 eV for $\text{BAs}_3$ as the concentration increases to $x = 4.50$. This penalty forces the OCV into the negative region (-0.26 V for $\text{BP}_3$ and -0.19 V for $\text{BAs}_3$). Because negative operating voltages cause dangerous lithium metal buildup, the maximum safe lithium capacity is limited to $x_{\max} = 3.0$.

    The sodiation process similarly begins with high initial voltages 1.62 V for $\text{BP}_3$ and 1.49 V for $\text{BAs}_3$ at $x = 0.125$ driven by strong host–adatom binding. As sodiation progresses up to $x = 3.00$, the system balances binding affinity against electrostatic Na–Na repulsion by maintaining a dispersed single-layer coverage, resulting in a dominant OCV plateau at approximately 0.4 V. Beyond $x = 3.00$, a transition in the storage mechanism occurs as monolayer coverage reaches capacity, necessitating the formation of secondary and tertiary metallic layers. Because these outer Na layers interact primarily with underlying sodium atoms rather than the $\text{BX}_3$ substrate, the average adsorption energy experiences a step-down reduction. This shift from monolayer adsorption to multilayer metallic clustering drives the final OCV plateaus down to 0.11 V for $\text{BP}_3$ and 0.03 V for $\text{BAs}_3$ at their respective maximum capacities of $x = 15.00$ and $x = 12.00$.

    Crucially, the OCV remains strictly positive throughout the full evaluated sodiation range. The fact that the adsorption energy remains strictly negative even at extreme sodiation states confirms that the multilayer sodiated phases are more energetically favorable than bulk sodium metal, guaranteeing a safe, dendrite-free operating mechanism near 0 V. Integrating the voltage plateaus yields exceptionally low average operating potentials: 0.39 V and 0.35 V vs. Li/Li$^+$ for BP$_3$ and BAs$_3$, and 0.18 V and 0.15 V vs. Na/Na$^+$. These values compare favorably with other 2D anode candidates such as elemental borophene (0.466 V) \cite{zhang2016could}, SnSe$_2$ (0.66 V) \cite{butt2021snse2}, PC$_6$ (0.40 V) \cite{yang2023potential}, tetragonal BN (0.35 V) \cite{xiong2024tetragonal}, and graphene-like AlP$_3$ (0.28 V) \cite{wan2023graphene}. Such low average potentials are highly advantageous as they maximize the overall full-cell output voltage when coupled with commercial high-potential cathodes, maximizing energy density while effectively mitigating thermal runaway risks.

    \begin{table}[htbp]
\centering
\caption{Comparison of theoretical specific capacities for $\text{BP}_3$ and $\text{BAs}_3$ monolayers against various 2D anode materials and conventional graphite for lithium- and sodium-ion batteries.}
\label{tab:capacity_comparison}
\begin{tabular}{llcc}
\toprule
 Type & Anode Material & Capacity ($\text{mAh/g}$) & Reference \\
\midrule
\multirow{9}{*}{Li-Ion} 
& Silicene (multi-layer) & $4200$ & \cite{tritsaris2013adsorption} \\
& Borophene & $1239$--$1860$ & \cite{zhang2016could,jiang2016borophene} \\
& $\text{V}_2\text{N}$ MXene (metallic) & $925$ & \cite{liu2022two} \\
& \textbf{$\text{BP}_3$ Monolayer} & \textbf{$775$} & \textbf{This work} \\
& Pristine Graphene & $744$ & \cite{sato1994mechanism} \\
& $\text{MoS}_2$ (TMD) & $\approx 600$ & \cite{stephenson2013lithium} \\
& Phosphorene & $432.79$ & \cite{zhao2014potential} \\
& Conventional Graphite & $372$ & \cite{nitta2015li} \\
& \textbf{$\text{BAs}_3$ Monolayer} & \textbf{$341$} & \textbf{This work} \\
\midrule
\multirow{11}{*}{Na-Ion}
& \textbf{$\text{BP}_3$ Monolayer} & \textbf{$3875$} & \textbf{This work} \\
& $\text{BP}_3$ Monolayer & $2325$ & \cite{vu2026bp} \\
& Dirac $\text{TiC}$ & $2015$ & \cite{sufyan2024monolayer} \\
& $\text{B}_3\text{P}$ & $1691$ & \cite{abbas2020two} \\
& \textbf{$\text{BAs}_3$ Monolayer} & \textbf{$1365$} & \textbf{This work} \\
& $\text{SiC}_2$ & $1203$ & \cite{li2022novel} \\
& MoPC & $1157$ & \cite{vu2026first} \\
& $h$-BAs & $522$ & \cite{khossossi2019ab} \\
& Janus $\text{MoSSe}$ & $510$ & \cite{wang2019two} \\
& $\text{Mo}_2\text{CO}_2$ MXene & $379$ & \cite{li2024theoretical} \\
& $\text{Ti}_3\text{C}_2\text{T}_x$ MXene & $125$ & \cite{gentile2025ti3c2t} \\
\bottomrule
\end{tabular}
\end{table}

    To quantitatively assess the energy storage potential, the theoretical specific capacity ($C$) was calculated based on the fully lithiated and sodiated ground-state configurations using the relation:
    \begin{equation}
    C = \frac{x_{\max} z F}{M}\label{eq:capacity},
    \end{equation}
    where $x$ is defined as the concentration ratio of adsorbed alkali atoms relative to the pnictogen atoms in the host matrix ($x = N_{\text{adatom}} / N_{\text{X}}$), and $x_{\max}$ represents the maximum stable stoichiometry of adsorbed alkali atoms per formula unit ($3.0$ for Li; up to $15.0$ and $12.0$ for Na on $\text{BP}_3$ and $\text{BAs}_3$, respectively). The term $z = 1$ is the valence charge number of the intercalant ion, $F = 26801\text{ mAh/mol}$ is the Faraday constant, and $M$ is the molar mass of the pristine $\text{BX}_3$ host framework ($M = M_{\text{B}} + 3M_{\text{X}}$). For the $2 \times 2 \times 1$ supercell model containing 8 boron atoms and 24 pnictogen atoms ($\text{B}_8\text{X}_{24}$), $x = 1/8$ ($0.125$) corresponds to a single adsorbed alkali atom. The individual atomic masses utilized are $10.81\text{ g/mol}$ for $\text{B}$, $30.97\text{ g/mol}$ for $\text{P}$, $74.92\text{ g/mol}$ for $\text{As}$, $6.94\text{ g/mol}$ for $\text{Li}$, and $22.99\text{ g/mol}$ for $\text{Na}$.

    For lithium-ion storage, the terminal capacity limit of $x = 3.0$ corresponds to a high theoretical gravimetric capacity of 775 mAh/g for the BP$_3$ monolayer. As summarized in Table~\ref{tab:capacity_comparison}, this value significantly outperforms conventional graphite (372 mAh/g) \cite{nitta2015li} and surpasses several prominent 2D anode candidates, including pristine graphene (744 mAh/g) \cite{sato1994mechanism}, phosphorene (432.79 mAh/g) \cite{zhao2014potential}, and typical transition metal dichalcogenides such as MoS$_2$ ($\approx$ 600 mAh/g) \cite{stephenson2013lithium}. While lower than the ultra-high capacities of borophene (1239–1860 mAh/g) \cite{zhang2016could,jiang2016borophene}, metallic V$_2$N MXenes (925 mAh/g) \cite{liu2022two}, and multi-layer silicene (4200 mAh/g) \cite{tritsaris2013adsorption}, the BP$_3$ framework circumvents the severe structural strain and restacking issues associated with extreme lithiation limits. Due to the heavier atomic mass of arsenic, the BAs$_3$ monolayer yields a reduced Li-storage capacity of 341 mAh/g.

    For sodium-ion storage, the multi-layer capability allows the BP$_3$ monolayer to reach an ultrahigh theoretical specific capacity of 3875 mAh/g, while the heavier BAs$_3$ monolayer delivers a substantial 1365 mAh/g. This performance vastly exceeds typical MXenes (e.g., Ti$_3$C$_2$T$_x$ at 125 mAh/g) \cite{gentile2025ti3c2t}, Mo$_2$CO$_2$ (379 mAh/g) \cite{li2024theoretical}, Janus MoSSe (510 mAh/g) \cite{wang2019two}, and the related h-BAs monolayer (522 mAh/g) \cite{khossossi2019ab}. Remarkably, it competes with or eclipses premier high-capacity anomalies in the literature, including SiC$_2$ (1203 mAh/g) \cite{li2022novel}, the phosphorus-based B$_3$P monolayer (1691 mAh/g) \cite{abbas2020two}, previous BP$_3$ SIB calculations (2325 mAh/g) \cite{vu2026bp}, and Dirac TiC (2015 mAh/g) \cite{sufyan2024monolayer}. This exceptional dual-ion storage capability originates from the unique structural topology of the puckered BX$_3$ lattice, which can seamlessly accommodate dense, multi-layer alkali metal configurations symmetrically on both sides of the sheet.

    \begin{figure*}[ht]
        \centering
        \includegraphics[width=18cm]{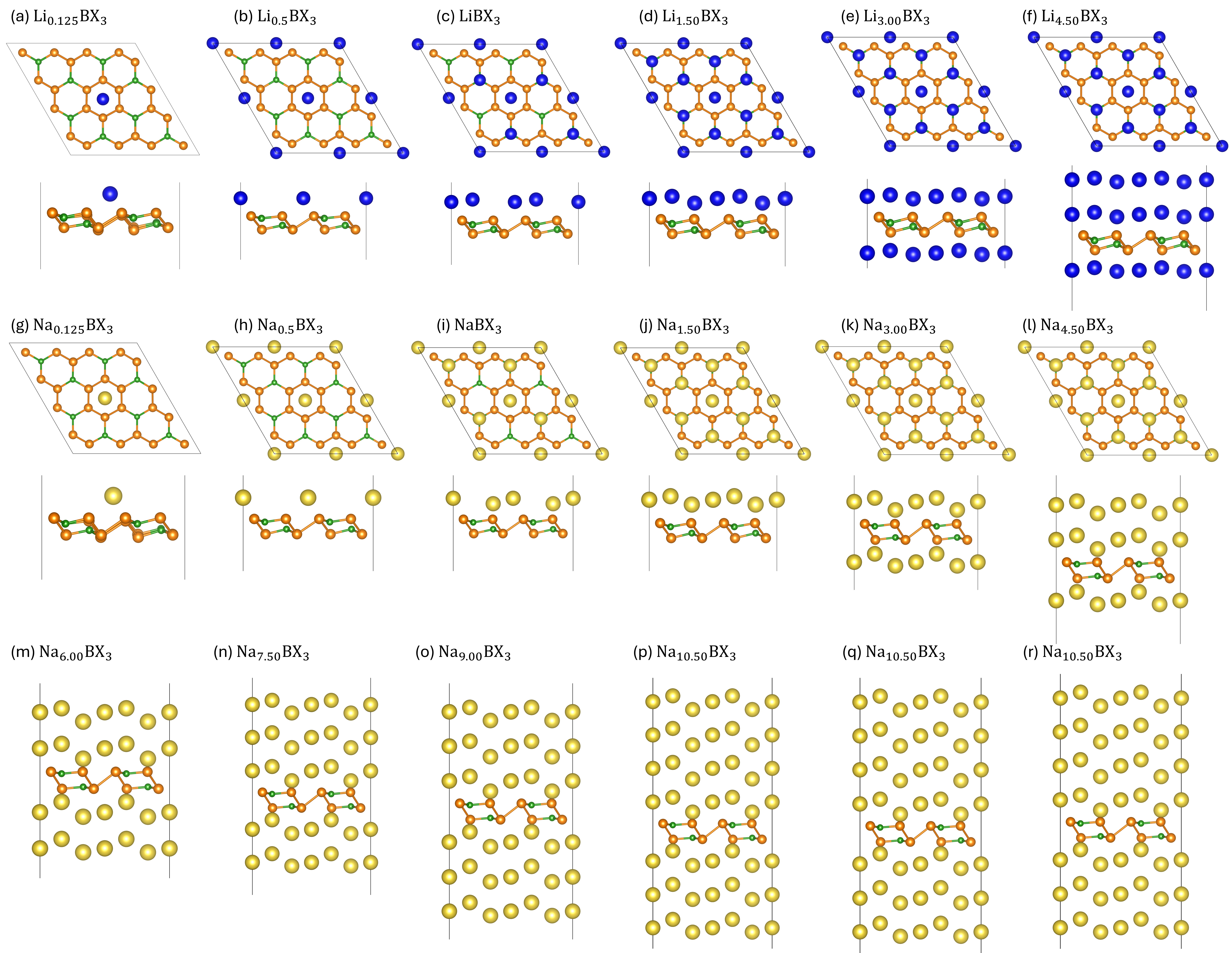}
        \caption{Top and side views of the optimized geometric configurations during progressive alkali metal adsorption on the $\text{BX}_3$ ($\text{X} = \text{P, As}$) monolayers. The panels illustrate various concentration levels (($x = N_{\text{adatom}}/N_{\text{X}}$, where $N_{\text{X}} = 8$)) explicitly denoted for both the lithiation ($\text{Li}_x\text{BX}_3$) and sodiation ($\text{Na}_x\text{BX}_3$) pathways. The structural evolution ranges from the initial dilute adsorption state at (a) and (g) $x = 0.125$ up to dense multi-layer metallic clustering at (f) $x = 4.50$ for Li and (r) $x = 10.50$ for Na. The green, orange, blue, and yellow spheres represent boron (B), pnictogen ($\text{X} = \text{P, As}$), lithium ($\text{Li}$), and sodium ($\text{Na}$) atoms, respectively.}
        \label{fig:charging_pathway}
    \end{figure*}

    To understand the stepped drops in the OCV profiles, we tracked the step-by-step filling sequence of alkali metals on the $\text{BX}_3$ monolayers (Fig.~\ref{fig:charging_pathway}). Isolated adatom energy calculations show that the site stability hierarchy follows $\text{H}_3 \rightarrow \text{B}_1 \rightarrow \text{O}_1 \rightarrow \text{H}_4$ for Na on both hosts, whereas for Li it follows $\text{H}_3 \rightarrow \text{H}_4 \rightarrow \text{B}_1 \rightarrow \text{O}_1$ on $\text{BP}_3$ and $\text{H}_3 \rightarrow \text{B}_1 \rightarrow \text{H}_4 \rightarrow \text{O}_1$ on $\text{BAs}_3$ (Fig. 4b, c). However, because the adjacent $\text{H}_4$ and $\text{B}_1$ sites lie in extreme spatial proximity, simultaneous occupation causes severe local Coulombic repulsion. To prevent this inter-ionic penalty, both systems dynamically bypass clustering at $\text{B}_1$ sites, establishing distinct effective filling pathways across the hosts: $\text{H}_3 \rightarrow \text{H}_4 \rightarrow \text{O}_1$ for Li and $\text{H}_3 \rightarrow \text{O}_1 \rightarrow \text{H}_4$ for Na.

    This sequential filling directly drives the structural evolution and corresponding voltage plateaus. At dilute initial concentrations ($x = 0.125$), the first Li and Na ions fill the most stable hollow $\text{H}_3$ sites, producing strong binding and setting the high initial voltage plateau (Fig.~\ref{fig:charging_pathway}a, g). As loading increases to $x = 1.50$, adatoms populate the secondary domains ($\text{H}_4$ for Li and $\text{O}_1$ for Na) to complete single-sided coverage on one face of the sheet (Fig.~\ref{fig:charging_pathway}b–d for Li; h–j for Na). Further loading up to $x = 3.00$ symmetrically fills both faces of the monolayer with a dense, double-sided single layer (Fig.~\ref{fig:charging_pathway}e, k).

    Beyond double-sided single-layer saturation ($x > 3.00$), Li and Na adapt differently to their respective thermodynamic limits. For lithium, forced over-loading ($x = 4.50$, Fig.~\ref{fig:charging_pathway}f) drives ions into high-energy tertiary positions, triggering an abrupt drop into unsafe negative working potentials. Conversely, continued sodium insertion (Fig.~\ref{fig:charging_pathway}l–r) avoids this energetic penalty by transitioning into multi-layer metallic clusters symmetrically on both sides of the host, enabling substantially higher storage capacities while maintaining safe, positive operating voltages.

\subsection{Alkali Metal Ion Diffusion Kinetics}

    \begin{figure}[ht]
        \centering
        \includegraphics[width=8.7cm]{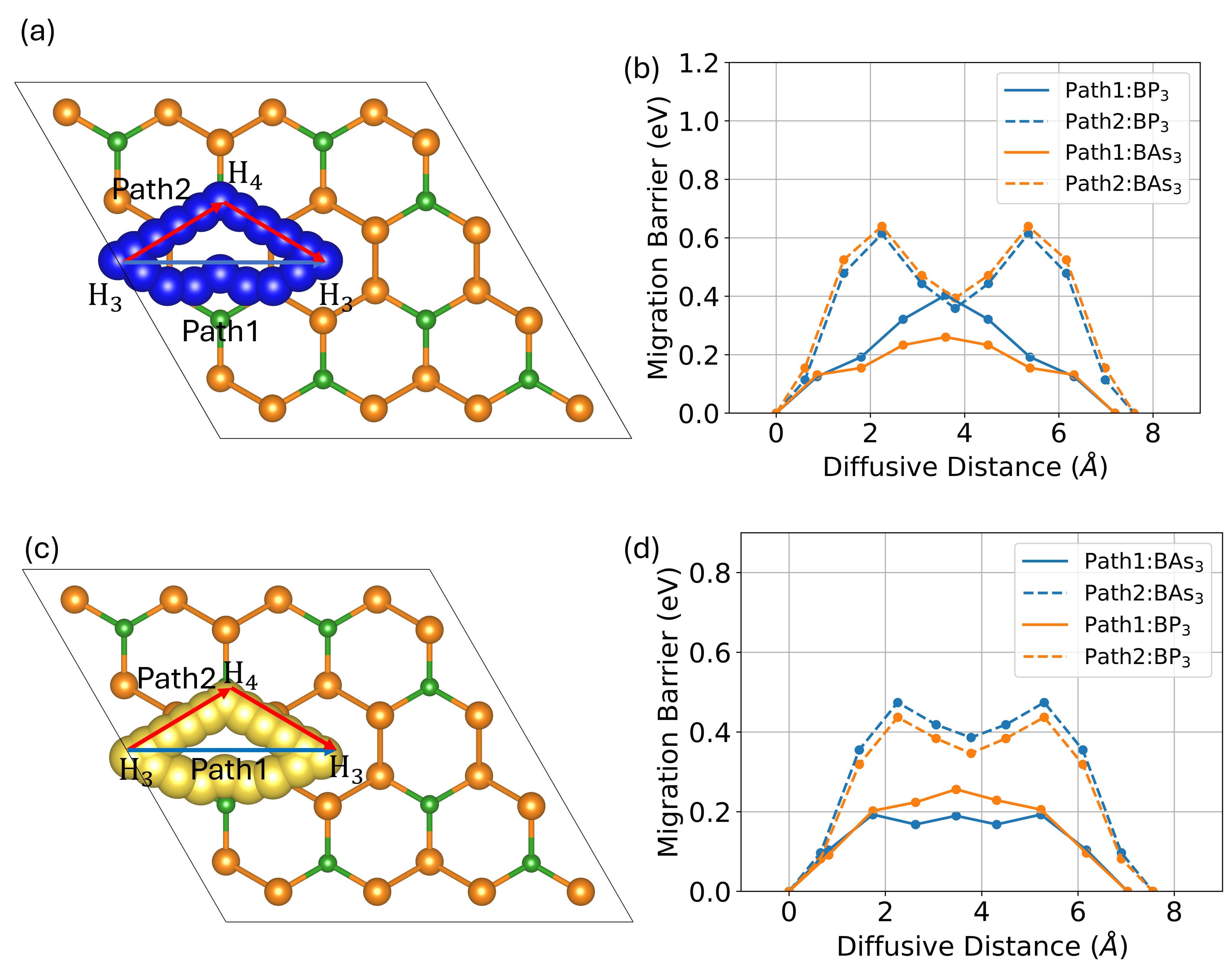}
        \caption{Alkali metal ion diffusion kinetics on the $\text{BX}_3$ ($\text{X} = \text{P, As}$) monolayers. (a, c) Top views of the migration pathways between adjacent thermodynamically stable $\text{H}_3$ sites for (a) lithium and (c) sodium. Path 1 (blue arrow) represents the direct diffusion route, while Path 2 (red arrows) represents an indirect diffusion route passing through an intermediate $\text{H}_4$ site. The superimposed blue and yellow spheres illustrate the nudged elastic band (NEB) image trajectories for Li and Na, respectively. (b, d) Calculated minimum energy pathways (MEPs) and migration barriers for (b) Li and (d) Na migration. Solid lines correspond to Path 1 and dashed lines correspond to Path 2 for the $\text{BP}_3$ and $\text{BAs}_3$ systems.}
        \label{fig:ci_neb}
    \end{figure}

    The rate capability of next-generation ion batteries relies heavily on rapid metal-ion diffusion kinetics across the anode surface. To evaluate ion mobility, activation barriers and migration pathways were investigated using the climbing image nudged elastic band (CI-NEB) method in a $2 \times 2 \times 1$ supercell \cite{henkelman2000climbing}. Minimum energy pathways (MEPs) were determined using 9 interpolated images for Path 1 and 11 images for Path 2. Because the $\text{H}_3$ hollow site represents the deepest thermodynamic energy minimum for isolated Li and Na adsorption, macroscopic transport is governed by inter-site $\text{H}_3 \rightarrow \text{H}_3$ hopping. We evaluated two routes (Fig.~\ref{fig:ci_neb}a, c): a direct linear pathway ($\text{H}_3 \rightarrow \text{H}_3$, Path 1) and an indirect pathway passing through an intermediate hollow $\text{H}_4$ site ($\text{H}_3 \rightarrow \text{H}_4 \rightarrow \text{H}_3$, Path 2). The resulting energy profiles and activation barriers ($\Delta E$) are presented in Fig.~\ref{fig:ci_neb}(b, d).

    Path 1 is unambiguously the preferred diffusion channel in all evaluated systems due to the smoother energetic topology along the direct vector and the avoidance of the less favorable $\text{H}_4$ site. For Li diffusion, Path 1 yields smooth, single-barrier profiles of 0.40 eV ($\text{BP}_3$) and 0.26 eV ($\text{BAs}_3$), whereas Path 2 imposes double-peaked barriers of 0.61 eV and 0.64 eV, respectively. For Na diffusion, Path 1 exhibits exceptionally low barriers of 0.26 eV ($\text{BP}_3$) and 0.19 eV ($\text{BAs}_3$), compared to 0.44 eV and 0.47 eV for Path 2. Notably, the Na Path 2 profile exhibits a local minimum at the $\text{H}_4$ site, a feature absent in previous computational evaluations \cite{vu2026bp}. Comparing hosts, $\text{BAs}_3$ exhibits intrinsically faster kinetics than $\text{BP}_3$ for both Li (0.26 eV vs. 0.40 eV) and Na (0.19 eV vs. 0.26 eV) owing to its larger lattice constant and wider diffusion channels. These low Path 1 barriers (all $< 0.50\text{ eV}$) predict rapid room-temperature transport via the Arrhenius relation. For Li storage, the 0.26–0.40 eV barriers are highly competitive with graphite ($\approx 0.40\text{ eV}$) \cite{nitta2015li} and $\text{MoS}_2$ ($\approx 0.25\text{ eV}$) \cite{stephenson2013lithium}, while remaining higher than elemental borophene ($< 0.01\text{ eV}$) \cite{zhang2016could} or anisotropic phosphorene (0.09 eV) \cite{zhao2014potential}. For Na storage, the 0.19–0.26 eV barriers indicate ultra-fast kinetics that compare favorably against reported 2D anodes including $\text{BP}_2$ (0.03 eV) \cite{ye2021metallic}, MoPC (0.06 eV) \cite{vu2026first}, Janus WSSe (0.07 eV) \cite{ahmad2023first}, $\text{SiP}_2$ (0.11 eV) \cite{wang2024sip2}, SnC (0.17 eV) \cite{butt2021monolayer}, BSi (0.24 eV) \cite{butt2021monolayer}, and $\text{BC}_3\text{N}_3$ (0.73 eV) \cite{xia2023boron}. These findings confirm that both architectures fulfill the prerequisites for high-power-density and rapid-charging applications.

\subsection{Alkali Metal Adsorption Configurations}
    \begin{figure}[ht]
        \centering
        \includegraphics[width=8.6cm]{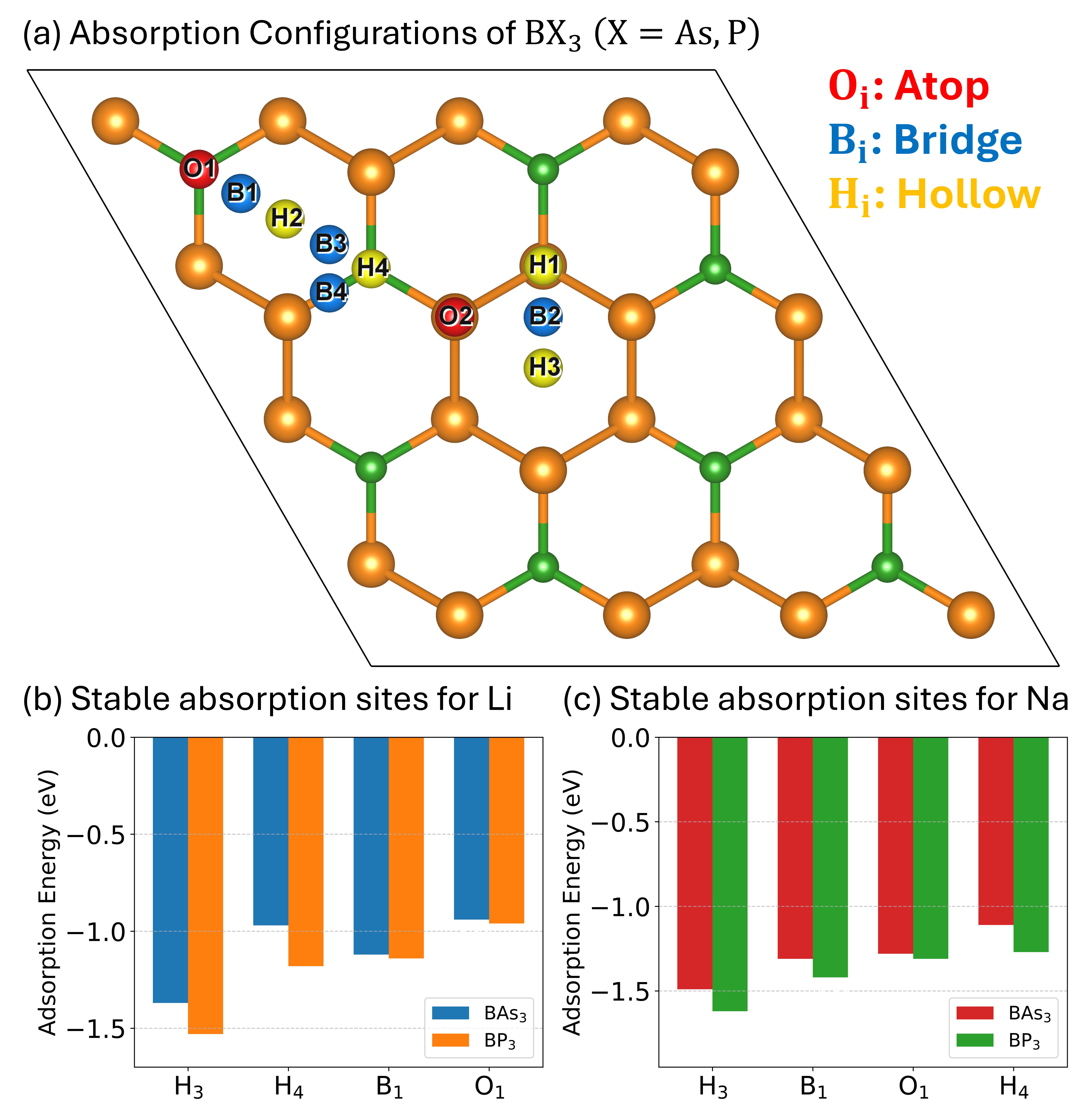}
        \caption{Alkali metal adsorption characteristics on the $\text{BX}_3$ ($\text{X} = \text{P, As}$) monolayers. (a) Top view of the pristine $\text{BX}_3$ lattice illustrating the considered high-symmetry initial adsorption sites: atop ($\text{O}_i$, red), bridge ($\text{B}_i$, blue), and hollow ($\text{H}_i$, yellow). The green and orange spheres represent the boron (B) and pnictogen ($\text{X} = \text{P, As}$) atoms, respectively. Calculated adsorption energies (in eV) of the four most thermodynamically stable configurations after full structural relaxation for (b) lithium ($\text{Li}$) adsorption on $\text{BAs}_3$ (blue bars) and $\text{BP}_3$ (orange bars), and (c) sodium ($\text{Na}$) adsorption on $\text{BAs}_3$ (red bars) and $\text{BP}_3$ (green bars).}
        \label{fig:ab_sites}
    \end{figure}

    \begin{figure*}[ht]
        \centering
        \includegraphics[width=18cm]{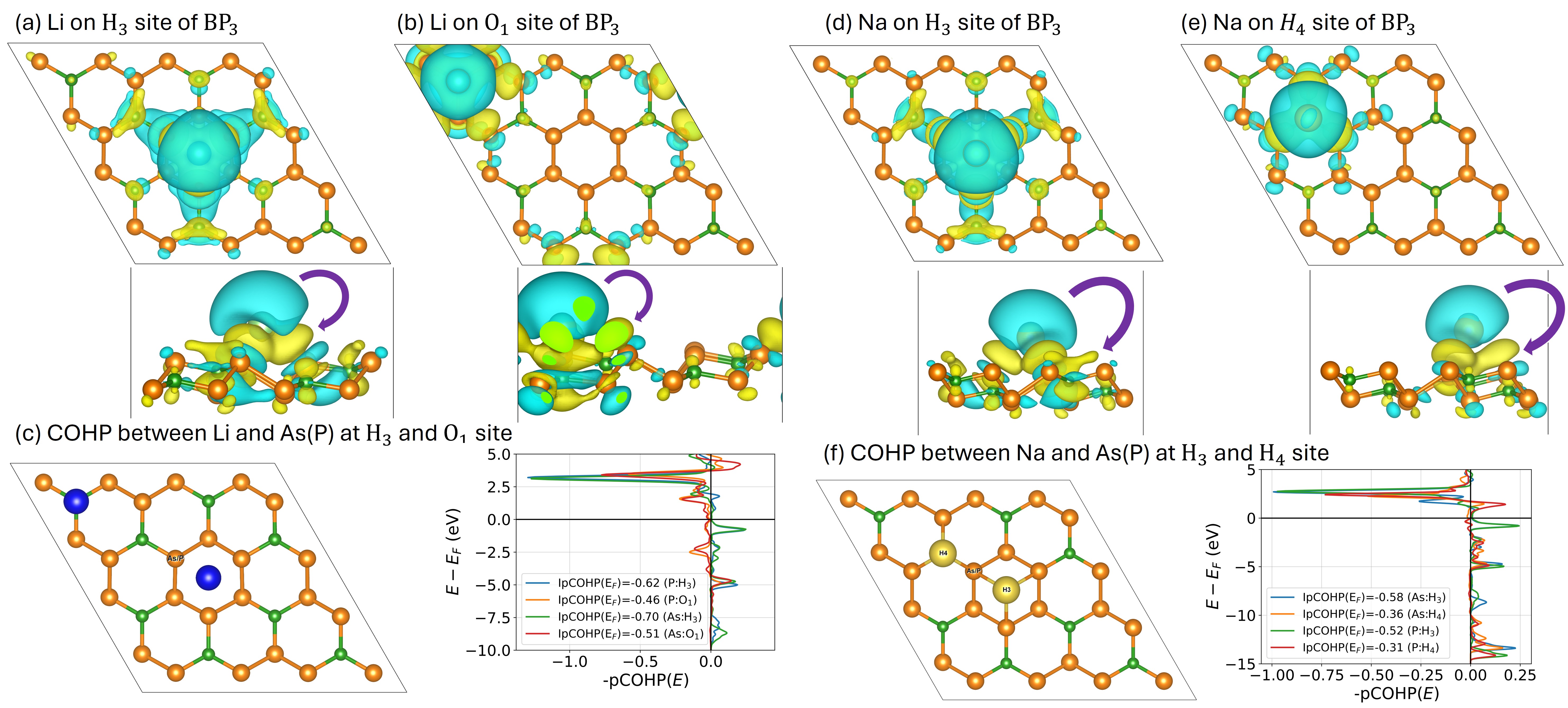}
        \caption{Electronic structure analysis and chemical bonding characteristics of alkali metal adsorption on the $\text{BX}_3$ ($\text{X} = \text{P, As}$) monolayers. Top and side views of the charge density difference (CDD) isosurfaces for Li adsorption on $\text{BP}_3$ at the (a) $\text{H}_3$ and (b) $\text{O}_1$ sites, and Na adsorption on $\text{BP}_3$ at the (d) $\text{H}_3$ and (e) $\text{H}_4$ sites. The cyan and yellow regions represent electron accumulation and depletion, respectively. (c, f) Schematic representations of the relevant high-symmetry adsorption sites alongside the calculated projected Crystal Orbital Hamilton Population ($-\text{pCOHP}$) profiles for (c) $\text{Li–P}$ and $\text{Li–As}$ interactions at the $\text{H}_3$ and $\text{O}_1$ sites, and (f) $\text{Na–P}$ and $\text{Na–As}$ interactions at the $\text{H}_3$ and $\text{H}_4$ sites. The inset legends provide the integrated COHP ($\text{IpCOHP}$) values at the Fermi level ($E_F$), quantifying the relative covalent bond strengths.}
        \label{fig:bader_cdd_cohp}
    \end{figure*}

    \begin{table}[htbp]
    \centering
    \caption{Convergence pathways from initial high-symmetry adsorption sites to final relaxed configurations and corresponding adsorption energies ($E_{\mathrm{ad}}$) for $\mathrm{Li}$ and $\mathrm{Na}$ atoms on $\mathrm{BP}_3$ and $\mathrm{BAs}_3$ monolayers.}
    \label{tab:convergence_energetics}
    \small
    \begin{tabular}{cc cc cc}
    \hline\hline
    \toprule
    \multirow{2}{*}{Ions} & \multirow{2}{*}{Final Site} & \multicolumn{2}{c}{$\mathrm{BP}_3$} & \multicolumn{2}{c}{$\mathrm{BAs}_3$} \\
    \cmidrule(lr){3-4} \cmidrule(lr){5-6}
    & & Initial Sites & $E_{\mathrm{ad}}$ (eV) & Initial Sites & $E_{\mathrm{ad}}$ (eV) \\
    \hline
    \multirow{4}{*}{ Li}
    & $\mathrm{H}_3$ & $\mathrm{O}_2, \mathrm{H}_1, \mathrm{H}_3, \mathrm{B}_2$ & -1.53 & $\mathrm{O}_2, \mathrm{H}_1, \mathrm{H}_3, \mathrm{B}_2$ & -1.37 \\
    & $\mathrm{H}_4$ & $\mathrm{H}_4, \mathrm{B}_3, \mathrm{B}_4$ & -1.18 & $\mathrm{H}_4, \mathrm{B}_3, \mathrm{B}_4$ & -0.97 \\
    & $\mathrm{B}_1$ & $\mathrm{H}_2, \mathrm{B}_1$ & -1.14 & $\mathrm{H}_2, \mathrm{B}_1$ & -1.12 \\
    & $\mathrm{O}_1$ & $\mathrm{O}_1$ & -0.96 & $\mathrm{O}_1$ & -0.94 \\
    \hline
    \multirow{4}{*}{ Na}
    & $\mathrm{H}_3$ & $\mathrm{H}_1, \mathrm{B}_2, \mathrm{O}_2$ & -1.62 & $\mathrm{H}_3, \mathrm{H}_1, \mathrm{B}_2, \mathrm{O}_2$ & -1.49 \\
    & $\mathrm{B}_1$ & $\mathrm{B}_1, \mathrm{B}_3, \mathrm{H}_2$ & -1.42 & $\mathrm{B}_1, \mathrm{B}_3, \mathrm{H}_2$ & -1.31 \\
    & $\mathrm{O}_1$ & $\mathrm{O}_1$ & -1.31 & $\mathrm{O}_1$ & -1.28 \\
    & $\mathrm{H}_4$ & $\mathrm{B}_4$ & -1.27 & $\mathrm{H}_4, \mathrm{B}_4$ & -1.11 \\
    \bottomrule
    \hline\hline
    \end{tabular}
    \end{table}

    Prior to evaluating the electrochemical performance of the $\text{BX}_3$ ($\text{X} = \text{P, As}$) systems, the geometric structures of the pristine monolayers were fully optimized. The $\text{BAs}_3$ monolayer exhibits an optimized in-plane lattice constant of $a = 6.98$ \AA\ and an intrinsic layer thickness of 1.39 \AA. In comparison, the $\text{BP}_3$ monolayer features a slightly more compact structure, with a lattice constant of $a = 6.50$ \AA\ and a thickness of 1.16 \AA. To eliminate any spurious interactions between periodic images along the out-of-plane direction, a large unit cell parameter of $c = 40.0$ \AA\ was employed. This dimension ensures a substantial vacuum space of at least 20 \AA, which is more than sufficient to accommodate the volumetric expansion of the systems even at their maximum theoretical multi-layer sodiation limits.

    To identify the optimal alkali metal ($\text{Li}$ and $\text{Na}$) storage mechanisms, we systematically investigated a comprehensive set of high-symmetry initial adsorption positions on both $\text{BX}_3$ surfaces. As illustrated in Fig.~\ref{fig:ab_sites}a, these are categorized into atop ($\text{O}_i$), bridge ($\text{B}_i$), and hollow ($\text{H}_i$) sites \cite{vu2026bp}. Following unconstrained structural relaxation, the various initial placements exhibited dynamic surface migration, ultimately converging into four unique and stable adsorption configurations for both materials: $\text{H}_3$, $\text{H}_4$, $\text{B}_1$, and $\text{O}_1$. As detailed in Table~\ref{tab:convergence_energetics}, the structural convergence pathways demonstrate a strong tendency for both Li and Na adatoms to migrate toward specific local energy minima. For example, initial adatom placements at the $\text{H}_1$, $\text{B}_2$, and $\text{O}_2$ sites consistently relaxed into the highly stable $\text{H}_3$ hollow site across all evaluated systems. Similarly, initial placements near the $\text{B}_4$ position consistently converged to the $\text{H}_4$ site, while the $\text{O}_1$ atop site remained stationary, retaining its initial configuration.

    Evaluating the calculated adsorption energies ($E_{\mathrm{ad}}$) of these converged configurations summarized in Fig.~\ref{fig:ab_sites}(b, c) and Table~\ref{tab:convergence_energetics} reveals that the $\text{H}_3$ site is unambiguously the most thermodynamically stable anchoring point for both ions. It yields highly exothermic binding energies for Li ($-1.53\text{ eV}$ on $\text{BP}_3$; $-1.37\text{ eV}$ on $\text{BAs}_3$) and for Na ($-1.62\text{ eV}$ on $\text{BP}_3$; $-1.49\text{ eV}$ on $\text{BAs}_3$). Beyond the primary $\text{H}_3$ site, the relative stability ordering of the remaining sites varies slightly depending on the intercalant. For Na adsorption, both monolayers follow the exact same energetic hierarchy: $\text{H}_3 > \text{B}_1 > \text{O}_1 > \text{H}_4$. For Li adsorption, the ordering shifts to $\text{H}_3 > \text{H}_4 > \text{B}_1 > \text{O}_1$ on the $\text{BP}_3$ framework, and $\text{H}_3 > \text{B}_1 > \text{H}_4 > \text{O}_1$ on $\text{BAs}_3$. Despite these minor mechanistic variations, all converged sites exhibit uniformly negative adsorption energies. This confirms a robust chemical affinity between the alkali metals and the puckered $\text{BX}_3$ networks, with the $\text{BP}_3$ monolayer consistently demonstrating slightly stronger ion binding compared to its arsenic counterpart.

\subsection{Electronic Structure, Bonding Mechanisms, and Conductivity}

    To gain deeper insights into the underlying interaction mechanisms and chemical bonding nature between the adsorbed alkali atoms and the host systems ($\text{BP}_3$ and $\text{BAs}_3$), charge density difference (CDD), Bader charge, and Crystal Orbital Hamilton Population (COHP) analyses were conducted. The spatial electron redistribution upon adatom insertion was evaluated using the relation:
    \begin{equation}
        \Delta\rho = \rho_{\text{BX$_3$+metal}} - \rho_{\text{BX$_3$}} - \rho_{\text{metal}},
    \end{equation}
    where the terms represent the charge densities of the adsorbed monolayer, the pristine host, and the isolated atom, respectively. The resulting spatial charge redistributions upon Li adsorption at the prominent $\text{H}_3$ and $\text{O}_1$ sites, and Na adsorption at the $\text{H}_3$ and $\text{H}_4$ sites, are illustrated in Fig.~\ref{fig:bader_cdd_cohp}(a,b,d,e). A substantial charge transfer from the alkali atoms to the surrounding host network is clearly visible. The pronounced accumulation of electronic charge (cyan clouds) resides primarily on the adjacent pnictogen ($\text{P/As}$) atoms, accompanied by a noticeable charge depletion (yellow regions) tightly localized around the adatom.

    Quantitatively, Bader charge analysis confirms that both lithium and sodium act as strong electron donors, revealing a prominent ionic character in the bonding mechanism. When localized at the primary $\text{H}_3$ site, the Li adatom loses $0.89\,e$ in the $\text{BP}_3$ system and $0.88\,e$ in the $\text{BAs}_3$ system. Similarly, the Na atom transfers approximately $0.84\,e$ to $\text{BP}_3$ and $0.83\,e$ to $\text{BAs}_3$ at the same site. Interestingly, this electron donation increases slightly when the ions occupy secondary sites; Li loses $0.92\,e$ ($\text{BP}_3$) and $0.91\,e$ ($\text{BAs}_3$) at the $\text{O}_1$ site, while Na donates $0.88\,e$ ($\text{BP}_3$) and $0.87\,e$ ($\text{BAs}_3$) at the $\text{H}_4$ site. From a purely electrostatic standpoint, this increased ionic character might imply that the secondary sites should exhibit the strongest binding energies. However, as established in the energetic analysis, the overall thermodynamic stability follows the opposite trend, indicating that ionic charge transfer alone is insufficient to explain the adsorption site preference, necessitating a deep evaluation of quantum mechanical orbital overlap.

    To resolve the specific covalent interactions between the adsorbates and the host atoms, we computed the projected Crystal Orbital Hamilton Population ($\text{pCOHP}$) implemented in the LOBSTER package. This method partitions the band-structure energy into bonding (positive $-\text{pCOHP}$) and antibonding (negative $-\text{pCOHP}$) contributions. As shown in Fig.~\ref{fig:bader_cdd_cohp}(c,f), the interactions between the alkali metals and the neighboring pnictogen atoms are heavily dominated by bonding states below the Fermi level ($E_F$). The integrated COHP ($\text{IpCOHP}$) up to $E_F$ serves as a quantitative metric for covalent bond strength where a more negative value denotes a stronger interaction. For lithiation, the $\text{Li–P}$ interaction on $\text{BP}_3$ yields an $\text{IpCOHP}$ of $-0.62\text{ eV}$ at the hollow $\text{H}_3$ site, which is noticeably stronger than the value of $-0.46\text{ eV}$ at the top $\text{O}_1$ site. The $\text{BAs}_3$ counterpart mirrors this trend, with the $\text{Li–As}$ interaction exhibiting a robust bonding value of $-0.70\text{ eV}$ at $\text{H}_3$ compared to $-0.51\text{ eV}$ at $\text{O}_1$. The sodiation data reveals an identical structural preference: the $\text{Na}$ atom at the $\text{H}_3$ site exhibits profound covalent binding ($-0.58\text{ eV}$ for $\text{Na–As}$ and $-0.52\text{ eV}$ for $\text{Na–P}$), while the $\text{H}_4$ site shows significantly weaker orbital hybridization ($-0.36\text{ eV}$ for $\text{Na–As}$ and $-0.31\text{ eV}$ for $\text{Na–P}$).

    Combining the ionic and covalent models provides a complete, synergistic picture of the adsorption mechanism. The highly electropositive alkali metal atoms bind to the surface through strong initial ionic interactions driven by near-complete valence electron donation, while their remaining electron clouds hybridize with the out-of-plane host orbitals to form stabilizing covalent bonds. The superior thermodynamic stability of the $\text{H}_3$ site is structurally driven: its geometric position within the primary puckered ring allows adatoms to achieve highly efficient, multi-center orbital overlap with the surrounding P or As atoms. This optimal coordination maximizes covalent binding strength heavily outweighing any minor deficit in ionic charge transfer relative to secondary sites and cements $\text{H}_3$ as the primary anchoring point across both hosts and chemistries.

    \begin{figure}[ht]
        \centering
        \includegraphics[width=8.7cm]{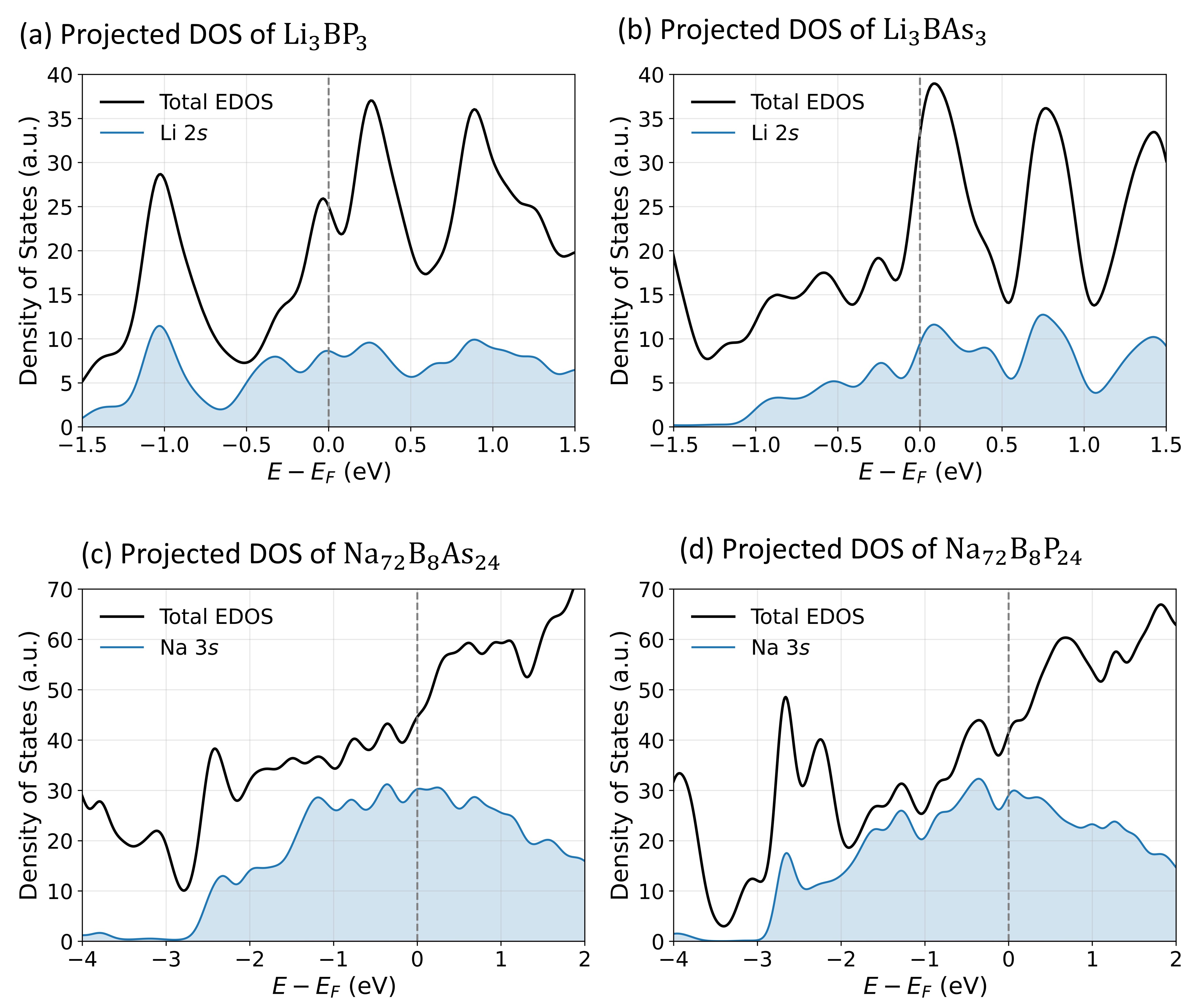}
        \caption{Calculated electronic properties of the $\text{BX}_3$ ($\text{X} = \text{P, As}$) frameworks at high alkali metal concentrations. Total electronic density of states (EDOS, solid black lines) and projected density of states (PDOS) for the corresponding alkali metal $s$ orbitals ($\text{Li}\ 2s$ or $\text{Na}\ 3s$, shaded blue profiles) for the lithiated (a) $\text{Li}_3\text{BP}_3$ and (b) $\text{Li}_3\text{BAs}_3$ phases, alongside the heavily sodiated supercell configurations of (c) $\text{Na}_{72}\text{B}_8\text{As}_{24}$ and (d) $\text{Na}_{72}\text{B}_8\text{P}_{24}$. The vertical dashed lines denote the Fermi level ($E_F = 0\text{ eV}$).}
        \label{fig:pdos}
    \end{figure}

    To evaluate electrical transport at maximum theoretical storage capacities, we computed the total and projected density of states (DOS) for the heavily intercalated configurations (Fig.~\ref{fig:pdos}). For both $\text{BP}_3$ and $\text{BAs}_3$, the total DOS at maximum lithiation ($x = 3.0$, Fig.~\ref{fig:pdos}a,b) and heavy sodiation ($\text{Na}_{72}\text{B}_8\text{X}_{24}$, Fig.~\ref{fig:pdos}c,d) reveals a continuous, finite distribution of electronic states crossing the Fermi level ($E_F$). This confirms that the intrinsic metallic character of the pristine host frameworks is preserved and enhanced upon dense alkali-metal insertion. This metallic nature is driven by prominent peaks emerging at and above $E_F$, heavily dominated by the valence $s$ orbitals of the intercalated ions ($\text{Li}\ 2s$ and $\text{Na}\ 3s$). Rather than remaining localized, the injected electrons populate the host conduction band and hybridize with host states, ensuring high intrinsic electrical conductivity during charge/discharge cycling. Maintaining continuous metallicity minimizes internal resistance and ohmic losses, facilitating the rapid electron transport required for fast-charging applications while reducing the need for conductive carbon additives. Additionally, uniform metallic charge delocalization helps screen local electrostatic repulsions, preventing structural degradation or phase transitions during long-term cycling.

\section*{Conclusions}
    In summary, we evaluated 2D $\text{BP}_3$ and $\text{BAs}_3$ monolayers as anode materials for Li- and Na-ion batteries using first-principles density functional theory. Structural and energetic analyses show that both hosts share identical relaxation patterns, with the hollow $\text{H}_3$ site serving as the most stable anchoring point for both alkali metals. This preference stems from a synergistic bonding mechanism combining substantial ionic charge transfer from the adatoms to the pnictogen framework with deep covalent orbital hybridization.

    CI-NEB calculations reveal that both systems strongly favor direct linear hopping between adjacent $\text{H}_3$ sites, bypassing indirect routes to avoid electrostatic repulsion. This pathway yields ultralow diffusion barriers: 0.40 eV ($\text{BP}_3$) and 0.26 eV ($\text{BAs}_3$) for Li, and 0.26 eV ($\text{BP}_3$) and 0.19 eV ($\text{BAs}_3$) for Na. These low barriers ensure ultrafast charge/discharge kinetics suitable for high-power applications. Electrochemical assessments confirm high storage capacities within safe operating voltage windows. For lithiation, stability is bounded at $\text{Li}_3\text{BX}_3$ to avoid negative potentials, giving low average working voltages (0.39 V for $\text{BP}_3$; 0.35 V for $\text{BAs}_3$) and theoretical capacities of 775 mAh/g ($\text{BP}_3$) and 341 mAh/g ($\text{BAs}_3$). For sodiation, multi-layer metallic clustering enables stoichiometry up to $\text{Na}_{72}\text{B}_8\text{X}_{24}$, delivering ultrahigh capacities of 3875 mAh/g ($\text{BP}_3$) and 1365 mAh/g ($\text{BAs}_3$) with average open-circuit voltages of 0.18 V and 0.15 V, respectively.

    Finally, density of states calculations confirm that intrinsic metallic conductivity is preserved across all intercalation levels for both ions. Electron injection from alkali $s$ orbitals into the host conduction band ensures efficient electron transport at maximum capacity, reducing the need for conductive additives and screening local electrostatic repulsions. Overall, these findings position $\text{BP}_3$ and $\text{BAs}_3$ monolayers as promising, high-performance anode candidates for next-generation energy storage.
\section*{Acknowledgments}
	This research project is supported by the Second Century Fund (C2F), Chulalongkorn University (Grant No. C2F PD-2320260067). High-performance computing facility in this Research is funded by Thailand Science research and Innovation Fund Chulalongkorn University (ST690022300001).

\bibliography{references}

\end{document}